\documentclass[
    ,final            % use final for the camera ready runs
    ,draft            % use draft while you are working on the paper
    ,numberedheadings % uncomment this option for numbered sections
  ]
  {aipproc}

\layoutstyle{6x9}

\newcommand{\gsim}{\hbox{\rlap{$^>$}$_\sim$}}

\begin{document}

\title{The Cannonball Model Of Long GRBs - Overview}


\author{Shlomo Dado}{address= {Physics Department, Technion, Haifa 32000, 
Israel}
}

\author{Arnon Dar}{address={Physics Department, Technion, Haifa 32000, 
Israel}
}

\begin{abstract} During the past ten years, the predictions of the 
cannonball (CB)  model of gamma ray bursts (GRBs) were repeatedly 
confronted with the mounting data from space- and ground-based 
observations of GRBs and their afterglows (AGs). The two underlying 
radiation mechanisms of the model, inverse Compton scattering (ICS) and 
synchrotron radiation (SR), provided an accurate description of the prompt 
and afterglow emission in all of the many well-sampled GRBs that were 
studied. Simple as they are, these two mechanisms and the burst 
environment were shown to generate the observed rich structure of the GRB 
light-curves at all observed frequencies and times.
\end{abstract}

\maketitle

%%%%%%%%%%%%%%%%%%%%%%%%%%%%%%%%%%%%%%%%%%%%
%% MAINMATTER
%%%%%%%%%%%%%%%%%%%%%%%%%%%%%%%%%%%%%%%%%%%%

\section{Introduction}

Two models have been used extensively to analyze Gamma ray bursts (GRBs) 
and their afterglows (AGs): the fireball (FB) model and the cannonball 
(CB) model. Despite their similar names, the two models were and still are 
entirely different, hence only one of them, if either, may provide a 
faithful physical description of GRBs. Until recently, the fireball (FB) 
model has been widely accepted as that model. However, the rich and 
accurate data that have been accumulated in recent years from space-
and ground-based observations have challenged this prevailing view on GRBs 
(for details see [1] and references therein):  Synchrotron radiation (SR) 
cannot explain simultaneously their prompt optical emission and their hard 
X-ray and $\gamma$-ray emission. The prompt hard X-ray and $\gamma$-ray 
pulses cannot be explained by SR from internal shocks generated by 
collisions between conical shells. Neither can SR explain their typical 
energy, spectrum, spectral evolution, pulse-shape, rapid spectral 
softening during their fast decay phase and the established correlations 
between various observables. Moreover, contrary to the predictions of the 
FB model, the broadband afterglows of GRBs are highly chromatic at early 
times, the brightest GRBs do not show jet breaks, and in canonical GRBs 
where breaks are present, they are usually chromatic and do not satisfy 
the `closure relations' expected from FB model `jet breaks'.

In spite of the above, the GRB community is not so critical and many 
authors believe that the GRB data require only some modifications of the 
standard FB model in order to accommodate the observations. Other authors 
simply ignore the failures of the FB model and continue the  
interpretation of the observations with the FB model taxonomy 
(`colliding conical shells', `internal and external shocks', `forward and 
reverse shocks', `continuous energy injection', `refreshed shocks') and 
parametrize the data with freely adopted formulae (e.g., `segmented power 
laws', 'exponential-to power-law components') which were never derived 
explicitly from any underlying physical assumptions (for recent examples, 
see, e.g. [2],[3],[4]).

The situation of the CB model is entirely different. In a series 
of publications, which were largely ignored by the rest of the GRB 
community, it was demonstrated repeatedly that the model correctly 
predicted the main observed properties of GRBs and reproduces successfully 
the diverse broad-band light-curves of both long GRBs ([1] and references 
therein) and short hard bursts (SHBs) [5]. Here we highlight this success 
of the CB model for long GRBs.

\section{The CB model}

In the CB model [6,7], {\it long-duration} GRBs and their AGs are produced 
by bipolar jets of highly relativistic plasmoids of ordinary matter [8,9] 
ejected in core-collapse supernova (SN) explosions [10]. 
It is hypothesized that an accretion disk 
or a torus is produced around the newly formed compact 
object, either by stellar material originally close to the surface of the 
imploding core and left behind by the explosion-generating outgoing shock, 
or by more distant stellar matter falling back after its passage [11]. As 
observed in microquasars, each time part of the accretion disk falls 
abruptly onto the compact object two jets of cannonballs (CBs) made of 
{\it ordinary-matter plasma} are emitted with large bulk-motion Lorentz 
factors in opposite directions along the rotation axis, fromwhere matter 
has already fallen back onto the compact object due to lack of rotational 
support. The prompt $\gamma$-ray and X-ray emission is dominated by 
inverse Compton scattering (ICS) of photons of the SN glory - 
light scattered and/or emitted by the pre-supernova wind blown  
from the progenitor star. The CBs' electrons Compton up-scatter the glory 
photons to $\gamma$ and X-ray energies and collimate them into a 
narrow beam  along the CBs' directions of motion.

A second mechanism besides ICS that generates radiation by a CB is 
synchrotron radiation (SR). The emitted CBs, which initially expand in 
their rest frame with the speed of sound in a relativistic plasma, merge 
within a short (observer) time into a few leading CBs. The beamed 
radiation of the CBs ionizes the wind/ejecta blown by the progenitor star 
and the interstellar medium (ISM) in front of them. In the CBs' rest 
frame, the ions continuously impinging on a CB generate within it a 
turbulent magnetic field, which is assumed to be in approximate energy 
equipartition with them. In this field the Fermi accelerated CB electrons 
and ISM intercepted electrons emit synchrotron radiation. The initial 
expansion of a CB produces a rapidly rising SR light-curve that begins to 
decline and traces the circumburst density of the pre-supernova 
wind/ejecta blown by the progenitor star into the roughly constant ISM 
density. Only when the CB has swept a mass comparable to its rest mass, 
does the continuous collision with the medium begin to decelerate it 
effectively. This results in a gradual steepening (break) of the SR 
light-curve into an asymptotic power-law decay.

\section{Prompt $\gamma$-ray and X-ray emission}

\subsection{The pulse shape and spectral evolution}
A GRB is a sum of pulses beginning at different times.
Let  $t$ denote the time after the beginning of such a pulse.
Its light-curve produced by inverse Compton scattering of
glory photons with a thin bremsstrahlung spectrum
by the bulk of the CB's electrons which are comoving with it,
is generally well approximated by [1]
\begin{equation}
E\, {d^2N_\gamma\over dt\,dE}(E,t)\approx
A\, {t^2/\Delta^2  \over(1+t^2/\Delta^2)^2}\,
e^{-E/E_p(t)} \approx e^{-E /E_p(0)}\, F(E\,t^2),
\label{ICSlc}
\end{equation}
where $A$ is a constant that depends on the CB's baryon number, on its
Lorentz and Doppler factors, on the density
of the glory light and on the GRB's redshift and distance, and
\begin{equation}
E_p(t)\approx  E_p(0)\, {t_p^2 \over t^2+t_p^2}\,,
\label{PeakE}
\end{equation}
with $t_p$ being the time when the ICS contribution
to $E\, d^2N_\gamma/ dE\, dt$ reaches its peak value.
It satisfies $E_p\!=\!E_p(t_p)$, where $E_p$ is the peak energy of 
the time-integrated spectrum. 
Thus, in the CB model, each ICS pulse in the GRB light-curve
is described by four parameters, $A,$
$\Delta(E),$  $E_p(0)$ and 
the beginning time of the pulse which is set to be 0.
Eq.~(\ref{ICSlc}), with $E_p$ given by Eq.~(\ref{PeakE}), describes well
the shape and the spectral evolution of GRB pulses and of early-time
X-ray flares. In particular, it correctly describes the rapid spectral
softening during the fast decline phase of the prompt emission in GRBs
and XRFs.  This is demonstrated in Figs.~1 and 2
for  the most energetic GRB with known redshift, 990123,
and for the faint XRF 060218.
\begin{figure}
  \includegraphics[height=.3\textheight]{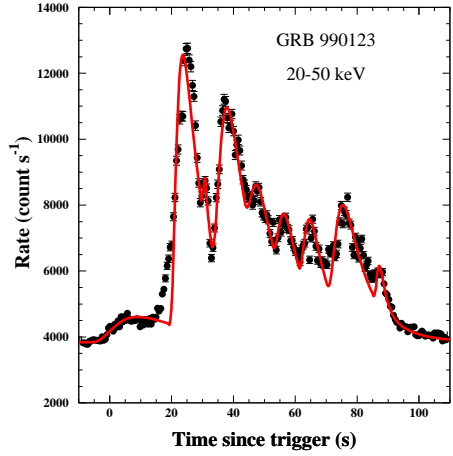}
  \caption{The BATSE light-curve of GRB 990123 and its CB model 
description [12]}
\end{figure}
\begin{figure}
  \includegraphics[height=.3\textheight]{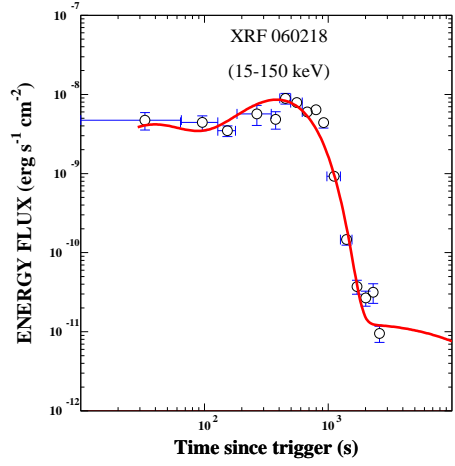}
  \caption{The Swift/BAT light-curve of XRF060218 and its CB model 
description [1]}
\end{figure}
If absorption in the CB is dominated by free-free absorption, then
$\Delta(E)\!\propto\! E^{-0.5}$, and then for $E\!<\!E_p$ 
the light-curve of an ICS peak 
is approximately a function of $E\,t^2$ (the `$Et^2$' law'),
with a peak
at $t\!=\!\Delta$, a full width at half maximum,
FWHM $\!\approx\!2\,\Delta$ and a rise time from half peak
value to peak value, ${\rm RT\!\approx\! 0.30\,FWHM}$. 

Note that the temporal decay of the energy flux 
of the prompt emission 
within an energy band, which follows
from Eqs.~(\ref{ICSlc}) and (\ref{PeakE}), is given approximately by,
\begin{equation}
\int_{E1}^{E2} E\, {d^2N_\gamma\over dt\,dE}(E,t)\, dE\approx 
A\, {E_p(t)\,\Delta^2\over t^2}\,
[e^{-E1/E_p(t)}-e^{-E2/ E_p(t)}].
\label{ICSlca}
\end{equation}
Thus, for the Swift XRT light-curves where 
$E1\!=\!0.3$ keV and $E2\!=\!10$ keV,
as long as $E_p(t) \!\gg\! E2\!\geq\!E1$, the energy 
flux decays like $t^{-2}$ until it is taken over 
by the SR afterglow, as demonstrated in
Fig.~3 for the bright GRB 061007.  
\begin{figure}
  \includegraphics[height=.3\textheight]{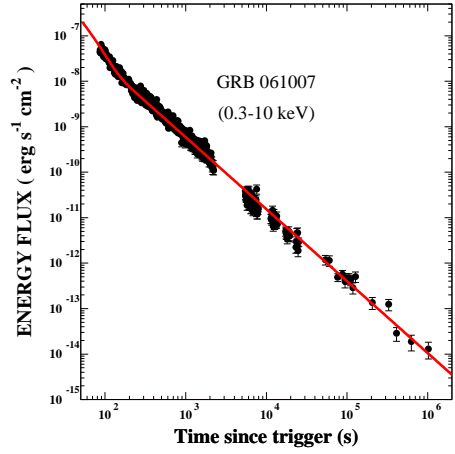}
  \caption{The Swift/XRT X-ray light-curve of GRB 061007,
and its CB model description [1]. The prompt emission 
decays like $t^{-2}$.}
\end{figure}
If $E1\!\ll\! E_p(t)$ but $E2\gsim E_p(t)$ the energy flux decays 
like $t^{-4},$ until it is taken over by the SR afterglow, as demonstrated 
in Fig.~4 for GRB 081221. 
\begin{figure}
  \includegraphics[height=.3\textheight]{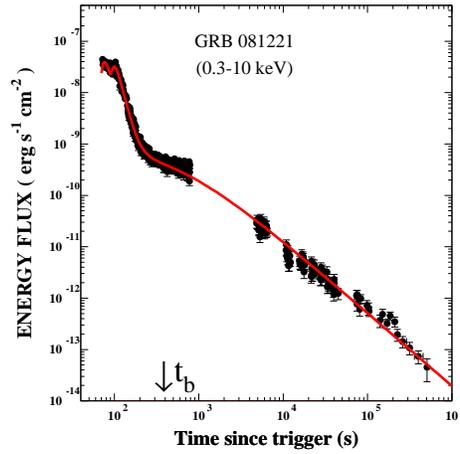}
  \caption{The Swift/XRT X-ray light-curve of GRB 081221
and its CB model description. The prompt emission
decays like $t^{-4}$.}
\end{figure} 
When $E1\gsim E_p(t)$ then the energy flux of the prompt emission 
decays like  $t^{-4}\, e^{-E\,t^2/2\, E_p\, t_p^2}$
until it is taken over by the SR afterglow,
as shown in Fig.~5.
\begin{figure}
  \includegraphics[height=.3\textheight]{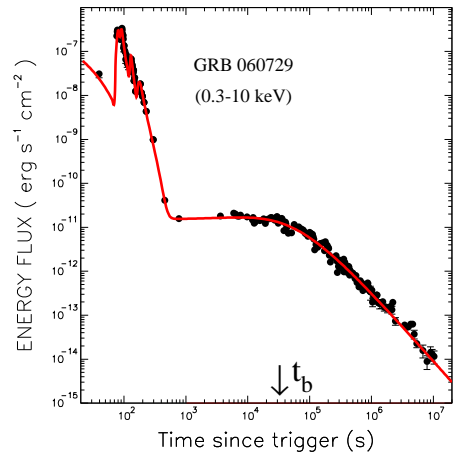}
  \caption{The Swift/XRT light-curve of GRB 060729 and its 
   CB model description [1]. The prompt emission pulses decay like $t^{-4}$
   times a Gaussian.}
\end{figure}.

\subsection{Polarization of the prompt emission}

The ICS of external unpolarized light by a highly relativistic
and narrowly collimated jet of CBs
results in a polarization of the hard X-ray and $\gamma$-ray emission
that is given approximately by [8,6],
\begin{equation}
\Pi(\theta,\gamma)\approx {2\;\theta^2\,\gamma^2\over
1+\theta^4\,\gamma^4},
\label{polSN}
\end{equation}
which, for the most probable viewing angles,
$\theta\approx 1/\gamma$, is of ${\cal{O}}(100\%)$.
The polarization of the prompt $\gamma$-ray emission has been measured in 
four GRBs [13-16] where a linear polarization $\Pi= (80\pm 20)\%$ 
in GRB 021206, $35\%\leq \Pi\leq 100\%$ in GRB 930131, 
$50\%\!\leq\! \Pi\! \leq 100\%$ in GRB 960924 and $\Pi\!=\! 98\% \! \pm\! 
33\%$ in GRB 041219A were obtained. Subsequent analyses of the case of GRB 
021206 by other groups questioned the result at the same level of 
significance [17,18], so that the 
degree of polarization of GRB 021206 remained uncertain.

\subsection{Correlations between the prompt emission observables}

The relativistic boosting and beaming of the glory photons 
by a CB yield
the relations [6,19]  $E_{iso}\propto \delta^3,$
$ (1+z)^2\, L_p\propto \delta^4,$ $(1+ z)\, E_p \propto \gamma\, \delta,$
and  $E_p\propto 1/\Delta,$
where $E_{iso}$ is the isotropic equivalent gamma ray energy, $L_p$ is 
the
peak isotropic equivalent luminosity, $\gamma$ is the bulk motion Lorentz
factor of a CB, and $\delta\! =\! 1/\gamma\, (1\!-\!\beta\, cos\theta)$ is
its Doppler factor with $\theta$ being the angle between the line of sight
to the CB and its direction of motion. For $\gamma^2 \gg 1$ and $\theta^2
\ll 1$, $\delta \approx 2\, \gamma/(1\!+\!\gamma^2\, \theta^2)$ to an
excellent approximation. The strong dependence of observables such as
$E_{iso}\,,$ $L_p$ $E_p$ and $\Delta t$ on $\gamma$ and $\delta$ and the
narrow distribution of $\theta$ around $1/\gamma$  result in
correlations among them [19] that are roughly represented  by
an average power-law,
\begin{equation}
(1 + z)\, E_p\propto E_{iso}^{0.50\pm 0.17}
             \propto [(1+z)^2\, L_p]^{0.375\pm0.125};~~~~
 \Delta t\propto 1/E_p\,.           
\label{correlations}
\end{equation}
The observed correlations between $(1 + z)\, E_p$ and $E_{iso}$
in GRBs with known redshift, $E_p$ and fluence 
are compared  in Fig.~6 with that predicted in the CB model [19, 20].
As shown in Fig.~6, the CB model correctly predicted the observed 
correlation, e.g., [21,22], between $(1+z)\,E_p$ and $E_{iso}$. 
\begin{figure}
  \includegraphics[height=.3\textheight]{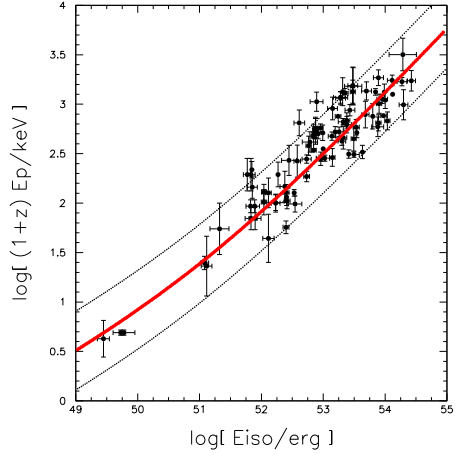}
  \caption{The observed correlation between 
$(1+z)\,E_p$ and $E_{iso}$ and that predicted by the CB model [19,20]
for long GRBs before 2009, with reliable $z$, $E_p$ and $E_{iso}$ 
values}
\end{figure}

\section{Synchrotron radiation}

The ISM ions continuously impinging on
a CB with a relative Lorentz factor $\gamma(t)$
generate
within it an equipartition turbulent magnetic field.
In this field the intercepted
electrons emit isotropic synchrotron radiation 
with  a characteristic frequency, $\nu'_b(t)$,
which is Doppler boosted
and collimated by its relativistic motion.
In the observer's frame:
\begin{equation}
\nu_b(t)\simeq  {\nu_0 \over 1+z}\,
{[\gamma(t)]^3\, \delta(t)\over 10^{12}}\,
\left[{n\over 10^{-2}\;\rm cm^3}\right]^{1/2}
{\rm Hz},
\label{nub}
\end{equation}
where $\nu_0\!\simeq\! 3.85\times 10^{16}\, {\rm Hz}\!\simeq\! 160$ eV/h. 
The spectral energy density of the SR
from a single CB at a luminosity distance $D_L$  is given by [7]:
\begin{equation}
F_\nu \simeq {\eta\,  \pi\, R^2\,n\, m_e\, c^3\,
\gamma(t)^2\, \delta(t)^4\, A(\nu,t)\,
\over 4\,\pi\, D_L^2\,\nu_b(t)}\;{p-2\over p-1}\;
\left[{\nu\over\nu_b(t)}\right]^{-1/2}\,
\left[1 + {\nu\over\nu_b(t)}\right]^{-(p-1)/2}\,,
\label{Fnu}
\end{equation}
where $p\!\sim\! 2.2$ is the typical spectral index
of the Fermi accelerated
electrons, $\eta\!\approx\!1$ is the fraction of the energy of the 
intercepted electrons that is synchrotron re-radiated, and $A(\nu,t)$
is the attenuation of photons of observed frequency $\nu$ along the line
of sight through the CB, the host galaxy (HG), the intergalactic medium
(IGM) and the Milky Way (MW):
\begin{equation}
A(\nu, t) = {\rm
exp[-\tau_\nu(CB)\!-\!\tau_\nu(HG)\!-\!\tau_\nu(IGM)\!-\!\tau_\nu(MW)].}
\label{attenuation}
\end{equation}
The opacity $\tau_\nu\rm (CB)$ at very early times, during the
fast-expansion phase of the CB, may strongly depend on time and frequency.
The opacity of the circumburst medium [$\tau_\nu\rm (HG)$ at early times]
is affected by the GRB and could also be $t$- and $\nu$-dependent.  The
opacities $\tau_\nu\rm (HG)$ and $\tau_\nu\rm (IGM)$ should be functions
of $t$ and $\nu$, for the line of sight to the CBs varies during the AG
observations, due to the hyperluminal motion of CBs.

The diverse behaviour of the broadband SR emission in
GRBs is well described by Eq.~(\ref{Fnu}) [1]. Due to lack of space we 
shall discuss only two of its important limits.

\subsection{The early-time SR}
The scattering of the wind's  particles by the CB 
stops its initial rapid expansion within a short time $t\!=\!t_{exp}$
[7]. During that time
both $\gamma$ and $\delta$ stay put at their initial values
and Eq.~(\ref{Fnu}) reduces 
to the early-time limit [1],
\begin{equation}
F_\nu \propto  {e^{-a/t}\,
t^{1-\beta} \over t^2+t_{exp}^2}\, \nu^{-\beta}\, ,
\label{SRP}
\end{equation}
where $\beta(t)\!=\!0.5$ for $\nu\!\ll\!\nu_b(t)$ and
$\beta(t)\!=\!p/2$ for $\nu\!\gg\!\nu_b(t)$.
Figs.~7 and 8 compare this prediction with the observed 
prompt optical emission in  GRBs 081203A [25] and 090102 [26], while 
Fig.~9
compares them for the brightest observed GRB, 080319B [3], where the 
prompt optical emission is resolved into contributions of 3 CBs
(or a single CB crossing a wind blown with interruptions). 
\begin{figure}
  \includegraphics[height=.3\textheight]{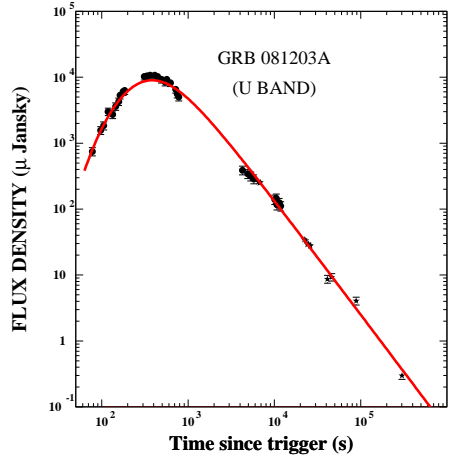}
  \caption{The $U$-band  light-curve
of GRB 080319B [3] and its CB model description}
\end{figure}
\begin{figure}
  \includegraphics[height=.3\textheight]{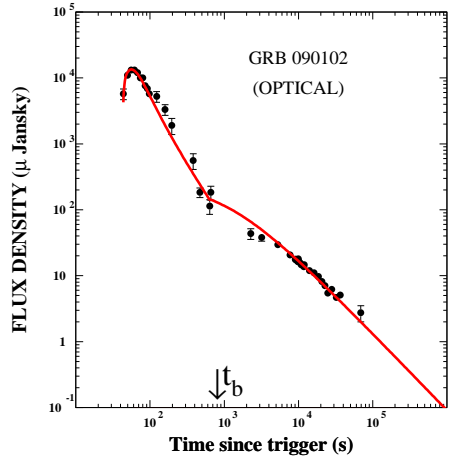}
  \caption{The $R$-band light-curve
   of GRB 090102 and its CB model description}
\end{figure}
\begin{figure}
  \includegraphics[height=.3\textheight]{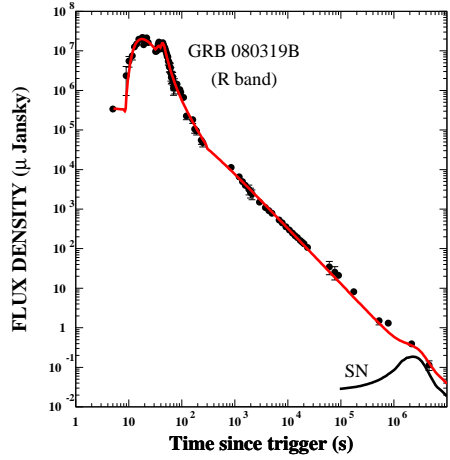}
  \caption{Comparison between the optical light-curve
           of GRB 080319B and that predicted
           by the CB model [12]}
\end{figure}

\subsection{SR during the CB's coasting phase}
The continuous collision with the medium decelerates the CB.
When the CB enters the constant density ISM, relativistic energy-momentum
conservation yields  the deceleration law 
([1] and references therein):
\begin{equation}
\gamma(t) = {\gamma_0\over \sqrt{[(1+\theta^2\,\gamma_0^2)^2 +t/t_0]^{1/2}
          - \theta^2\,\gamma_0^2}}\,,
\label{goft}
\end{equation}
with $t_0={(1\!+\!z)\, N_{_{\rm B}}/ 8\,c\, n\,\pi\, R^2\,\gamma_0^3}.$
As can be seen from Eq.~(\ref{goft}), $\gamma$  and $\delta$
change little as long as $t\!\ll\! t_b\!=\![1\!+\gamma_0^2\,
\theta^2]^2\,t_0\,, $ and Eq.~(\ref{Fnu}) yields the {\it `plateau'}
of canonical AGs.
For $t\!\gg\!t_b$, $\gamma$  and $\delta$ decrease like $t^{-1/4}\,.$
The transition $\gamma\!\sim\! \gamma_0$ to 
$\gamma\!\sim\! \gamma_0\,(t/t_0)^{-1/4}$
around $t_b$
induces a bend, the so-called `jet  break',
in the synchrotron AG 
from a plateau to an asymptotic power-law
$F_\nu \propto t^{-p/2-1/2}\,\nu^{-p/2}= t^{-\Gamma+1/2}\,
\nu^{-\Gamma+1}.$
In terms of the frequently used notation,
this asymptotic behaviour satisfies
$F_\nu(t)\propto t^{-\alpha}\,\nu^{-\beta}$
with $\alpha\!=
\!\beta\!+\!1/2\!=\Gamma\!-\!1/2\, .$
For a density 
$n\!\propto\! 1/r^2$, 
the asymptotic relation becomes
$\alpha\!=\!\beta+1\!=\!\Gamma,$
where $t$ is the time after the onset of the $n\!\propto\! 1/r^2$
density. These relations are well satisfied by the late-time 
power-law decay 
of canonical and non-canonical AGs and of late-time SR flares [1,23]. 

\subsection{Jet breaks and `missing breaks'}  
In terms of typical CB-model values of $\gamma_0$,
$R$, $N_{_{\rm B}}$ and $n$,
\begin{equation}
t_b= (1300\,{\rm s})\, [1+\gamma_0^2\, \theta^2]^2\,(1+z)
\left[{\gamma_0\over 10^3}\right]^{-3}\,
\left[{n\over 10^{-2}\, {\rm cm}^{-3}}\right]^{-1}
\left[{R\over 10^{14}\,{\rm cm}}\right]^{-2}
\left[{N_{_{\rm B}}\over 10^{50}}\right] \! .
\label{tbreak}
\end{equation}
Consequently, 
for a large density, a large $\gamma_0$ and
a small viewing angle, which correspond to large 
values of $E_{iso}$, $L_p$ and $E_p$, the break time $t_b$ 
becomes very small and may `hide' under the prompt ICS radiation, or occur 
too early to be seen by the Swift XRT [23]. In that case the 
X-ray AG measured by Swift XRT has 
the simple asymptotic power-law
decay, $F_\nu(t)\!\propto\! 
t^{\!-\!\beta_X\!-\!1/2}\,\nu^{\!-\!\beta_X},$
as was observed for several GRBs such as 050717, 061007, 071025, 080319B,
080804 and 081109 (see e.g., Fig.~3). 

\section{Break time - prompt emission correlations}
For a constant ISM density,  $t_b/(1+z)\!\propto\! 1/\gamma_0\delta_0^2$. 
Consequently, the favoured viewing angle 
$\theta\!\approx\!1/\gamma_0\,$, i.e., $\delta_0\approx \gamma_0\, ,$
and Eq.~(\ref{correlations}) imply correlations roughly represented by 
[23]  $t_b/(1\!+\!z)\!\propto\! 
E_{iso}^{-1}$, $t_b/(1\!+\!z)\!\propto\![(1\!+\!z)\,Ep]^{3/2}$, etc.,
while for large viewing angles the strong dependence on $\delta_0$
yields $t_b/(1\!+\!z)\!\propto\! E_{iso}^{-2/3}$ and
$t_b/(1\!+\!z)\!\propto\![(1\!+\!z)\,E_p]^2$, etc. These limits
can be well interpolated by formulae such as,
\begin{equation}
{t_b \over 1+z} \approx {2\,t_{b,eiso}\over (E_{iso}/E_0)^{2/3}+ 
E_{iso}/E_0} \sim t_{b,eiso}\, \left({E_{iso}\over E_0}\right)^{-0.83\pm 
0.17}.
\label{tbeiso}
\end{equation}
The predicted correlation between $t_b$ and $E_{iso}$
is compared in Fig.~10
to data on Swift GRBs prior to January 1, 2009 that 
have  a well measured $E_{iso}$ and a well sampled
X-ray light-curve.
\begin{figure}
  \includegraphics[height=.3\textheight]{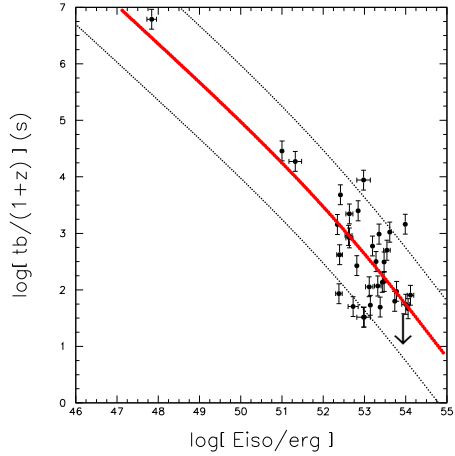}
  \caption{The observed correlation between
$t_b/(1+z)$ and $E_{iso}$ for long GRBs with well measured 
redshift and $E_{iso}$
and that predicted by the CB model [23]}
\end{figure}
The explanation for `missing AG breaks' is supported by the observed 
correlations between the break-time and the prompt emission observables 
[23].

\section{Flares}
In more than 50\% of the GRBs observed by Swift, the X-ray light-curve, 
during the prompt GRB and its early AG phase, shows flares superimposed on 
a smooth background. In the CB model, X-ray flares without an accompanying 
detectable $\gamma$-ray emission can be of two kinds. They can be ICS 
flares produced by CBs ejected with a relatively  small Lorentz factor 
and/or a large viewing angle. Such CBs may be ejected in accretion 
episodes both during the prompt GRB and in delayed accretion episodes onto 
the newly formed central object in core collapse SNe [11]. ICS flares 
satisfy the $E\,t^2$-law and exhibit a rapid softening during their fast 
decline phase that is well described by Eqs.~(\ref{ICSlc}) and 
(\ref{PeakE}). Often, during the rapidly decreasing phase of the prompt 
emission, there are `mini X-ray flares' that show this rapid spectral 
softening [24]. As the accretion material is consumed, one may expect the 
`engine' to have a few progressively-weakening dying pangs. Flares can 
also result from enhanced synchrotron emission during the passage of CBs 
through over-densities produced by mass ejections from the progenitor star 
or by interstellar winds [7]. Late flares seem to have the typical SR 
spectrum and spectral evolution induced by the dependence of $\nu_b$ on 
density.

\section{Conclusion}
The widely ignored CB model continues to be a remarkably successful 
model of GRBs.

\end{document}